\documentclass{aa}
\usepackage{amssymb,amsmath,natbib,graphicx,rotating}
\usepackage[format=hang]{caption,subfig}
\bibpunct{(}{)}{,}{a}{}{,}

\newcommand \degs  {$^{\circ}$}

\title{All-sky Galactic radiation at 45 MHz and spectral index between 45
  and 408 MHz} \author{A. E. Guzm\'an \inst{1}\and J. May \inst{1}\and
  H. Alvarez \inst{1}\and K. Maeda \inst{2}} \institute{Departamento de
  Astronom\'ia, Universidad de Chile, Casilla 36-D, Santiago, Chile \and
  Hyogo University of Health Sciences} \date{Received / Accepted }
\abstract{}{We study the Galactic large-scale synchrotron emission by
  generating a reliable all-sky spectral index map and temperature map at
  45 MHz.}{We use our observations, the published \mbox{all-sky} map at 408 MHz,
  and a bibliographical compilation to produce a map corrected for
  zero-level offset and extragalactic contribution.}{We present full sky
  maps of the Galactic emission at 45 MHz and the Galactic spectral index
  between 45 and 408 MHz with an angular resolution of 5$^\circ$.  The
  spectral index varies between 2.1 and 2.7, reaching values below 2.5 at
  low latitude because of thermal free-free absorption and its maximum in
  the zone next to the Northern Spur. }{}

\begin{document}

\authorrunning{Guzm\'an et al.}
\titlerunning{45 MHz all-sky map and 45-408 MHz spectral index.}
\maketitle 

\section{Introduction}
A way to describe some global features of our Galaxy is to perform radio
observations that span large areas of the sky. Below 1.4 GHz, there are 
 few  surveys with large  coverage 
and  all-sky maps such as
the 408 MHz maps of \citet{Haslam1981,Haslam1982} and the 1420 MHz maps
of \citet{Reich1982,Reich1986}, and \citet{Reich2001}.
The 45 MHz survey presented in this paper  covers  96\% of
 the sky. 

This frequency range probes mostly the synchrotron emission from
high energy electrons interacting with the large-scale magnetic field
of the Galaxy and the spectral information derived at these
frequencies is related to the spectral energy distribution of these
relativistic electrons. 
Spectral index maps are presented in \citet{Reich2004} 
between the frequencies  408, 1420, and 22800 MHz.
The subject has also received attention in its
application to the problem of adequately subtracting the foreground
radiation from the cosmic microwave background \citep{Oliveira2008}.

Important physical information can be extracted from the surveys and
the spectral index maps concerning Galactic structure, global
magnetic fields and relativistic electron distribution. In this paper,
we present an all-sky spectral index map by using our 45 MHz 
survey and the \mbox{408 MHz} \citep{Haslam1981,Haslam1982} all-sky
map, together with multi-frequency studies of six zones that allow us
to obtain zero-level corrections for both maps. In the Appendix,
we estimate  the extragalactic background non-thermal spectrum
based on a literature compilation.

 \section{The 45 MHz survey}
  The 45 MHz southern and northern sky maps \citep{45south,45north}
were combined in the all-sky map\footnote{The 45 MHz map can be downloaded from the website http://www.das.uchile.cl/survey45mhz.} shown in Fig. 1.  Missing data around
the north equatorial pole represents $\sim4\%$ of the whole sky.  The
southern map data were observed between 1982 and 1994 with an array of
528 E-W dipoles that produced an effective beam of
$4.6^\circ(\alpha)\times2.4^\circ(\delta)$ FWHM. A full description of
the instrument was given by  \citet{45ins} and an  experimental 
 determination of the beam in \citet{45insex}.
The northern data were taken in the periods of 1985-1989 and 1997-1999
with the MU radar array \citep{MUradara,MUradarb} at an angular
resolution of $3.6^\circ\times3.6^\circ $.  Table \ref{tabsurveys}
displays some of the principal characteristics of  southern and  northern
parts of the 45 MHz survey.

\begin{table*}
\caption{Parameters of the Northern and Southern surveys.}
\label{tabsurveys}
\centering
\begin{tabular}{ccccc}\hline\hline
Array&Coverage& Effective Area & Operating frequency&Resolution
(FWHM) \\ \hline Maip\'u & $-90^\circ\le\delta\le+19.1^\circ $ & 11200
m$^2$ & 45 MHz &$4.6^\circ(\alpha)\times2.4^\circ(\delta)$ \\ MUradar
&$+5^\circ\le\delta\le+65^\circ $ & 8300 m$^2$ & 46.5 MHz&
$3.6^\circ\times3.6^\circ $ \\ \hline
\end{tabular}
\end{table*}%

An important characteristic of the all-sky 45 MHz map is that it consists of only two surveys and uses only two transit instruments with similar
characteristics.\\ Some relevant features of this map are: 

  \begin{figure*}[h!]
\centering
\includegraphics[width=1.\textwidth ]{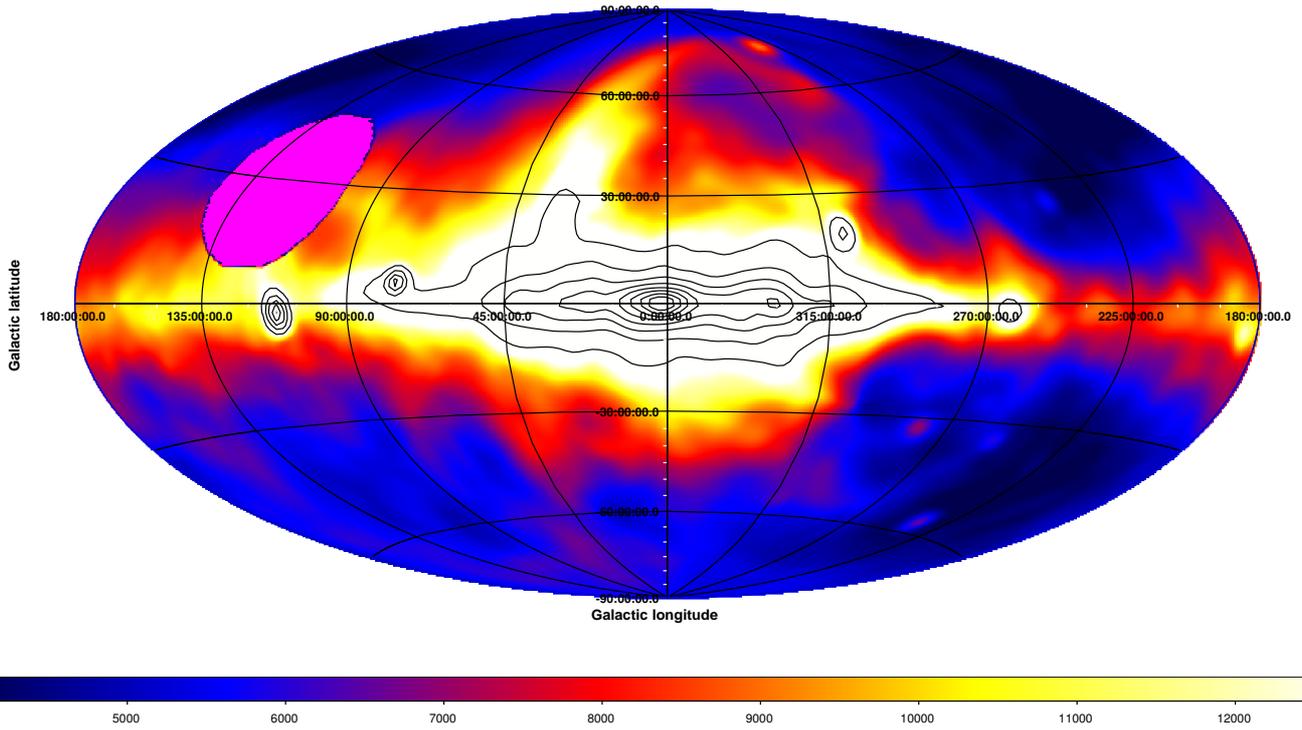}
\caption{Hammer-Aitoff projection of the 45 MHz full sky map. Eight contours are drawn  between 15000 and 60000 K. The map does not cover 
the  $\delta>+65^\circ$ zone.}\label{45sky}
\end{figure*}

 \begin{figure}
\centering
 \hspace*{-1.5 em}\includegraphics[width= .47 \textwidth, height=.43 \textwidth]{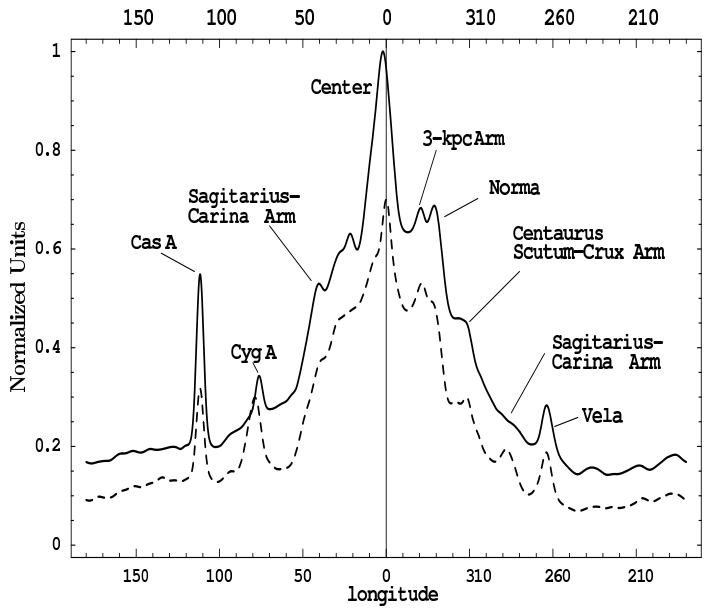}
\caption{Temperature averaged across latitude in the $|b|<10^\circ$
strip from 45 MHz (continuous) and 408 MHz (dashed) surveys in  arbitrarily
normalized units: T(45 MHz)/(54000 K) and T(408 MHz)/(300 K). The
"steps" mark the tangential directions to spiral arms.}\label{planes}
\end{figure}

\vspace*{-1 ex} \begin{enumerate}
 \item{Galactic plane: Most of the emission at 45 MHz comes from the plane
   of the Galaxy that has an overall maximum near the Galactic center.
   Figure \ref{planes} shows the plane profiles averaged inside the
   $|b|<10^\circ$ strip for the 45 and 408 MHz surveys, both convolved to
   the common $5^\circ\times5^\circ$ resolution.  We can see that most of
   the structure of the 408 MHz Galactic plane is reproduced in the 45 MHz
   plane.  This structure includes the strong maximum near the Galactic
   center (note that the 45 MHz map reaches its maximum at approximately
   $l=2^\circ$, as discussed in \citealt{Alvarez1997}), the ``steps'' in
   the emission produced by the spiral arms can be seen tangentially, and
   some strong sources exist near the plane.  {The 45 MHz map shows
     more conspicuously sources associated with non-thermal emission (such
     as the Galactic center and Cas A), compared to the 408 MHz map that
     shows strong thermal sources as well.}}
   
 \item{Northern spur: This feature forms an arc from
 $l\approx35^\circ,\,b\approx16^\circ$ passing through
 $l\approx0^\circ,\,b\approx75^\circ$ and ending near the position { of}
 \object{Virgo A}. { The hypothesis that this is a  nearby supernova remnant shell
is an old one  \citep{Hanbury1960}  and has become  
widely accepted (see \citealt{Wolleben2007} and references therein).}}
\item{Galactic and extragalactic sources: i) Galactic: SNR \object{Cas A}
($l=111.5^\circ\,b=-2.3^\circ $), \object{Vela SNR} ($l=263.9^\circ\,
b=-3.3^\circ$), and the direction of the Local arm towards Cygnus and
Vela.\\ ii) Extragalactic: Radio-galaxies such as \object{Cen A}
($l=309.5^\circ,\, b=19.4^\circ $) , \object{Cyg A}
($l=76.2^\circ,\,b=5.8^\circ $), \object{Virgo A}
($l={ 285}^\circ,\,b=74.3^\circ $), \object{Hydra A}
($l={ 243}^\circ,\,b=+25^\circ $), 
\object{Pictor A} ($l={ 252}^\circ,\,b=-34^\circ
$), and \object{Fornax A} ($l={ 240}^\circ,\,b=-56.3^\circ $). We can also
distinguish the Large Magellanic Cloud at
$l={ 280.2}^\circ,\,b=-32.8^\circ$.}

\item{Minimum temperature zones: There are two areas with minima of
  temperature, one in the southern Galaxy ($l\sim235^\circ,\,b\sim-45^\circ
  $, $T\sim3600$ K) and the other in the north ($l\sim192^\circ,\,
  b\sim48^\circ $, $T\sim 3800$ K).   It is remarkable that these   two points are rather close in Galactic longitude, at approximately the point where the Galactic plane reaches its minimum temperature compared to other longitudes ($T\sim8000$ K, $l\sim 225^\circ$, see Fig. \ref{planes}). It is in these
  directions that any extragalactic or instrumental feature should be most
  noticeable. These two zones are later discussed  in detail.}
 \end{enumerate}
 
\begin{table*}
\caption{Multifrequency data for the selected points. The asterisks indicate data not included in the  multifrequency fit.\label{tablaseiszonas}} 
\scriptsize
\begin{tabular}{r|cccccc|l}\hline \hline
 $\nu$ & Anti. & Calib.    & Minim.    & Minim.    & NGP& SGP & Ref. \\
\-[MHz]  &    [K]& Point [K] & North [K] & South [K] & [K]& [K] &      \\ \hline
1.6 &\ldots&\ldots&\ldots& $*\,6.5\text{E}6\pm 5.\text{E}5$ &\ldots&\ldots& 1  \\
 2.3 &\ldots&\ldots&\ldots&\ldots& $1.1\text{E}7\pm 1.\text{E}6$ & $1.8 \text{E}7\pm 1.\text{E}6$ & 2  \\
4.7 &\ldots&\ldots&\ldots& $*\,1.88\text{E}6\pm 4.\text{E}5$ &\ldots&\ldots& 3  \\
4.7 &\ldots&\ldots&\ldots& $*\,1.38\text{E}6\pm 1.\text{E}5$ &\ldots&\ldots& 4 \\
5.2 &\ldots&\ldots&\ldots&\ldots& $1.7\text{E}6\pm 4.\text{E}5$ & $1.3 \text{E}6\pm 3.\text{E}5$ & 5 \\
5.5 &\ldots&\ldots&\ldots& $*\,7.5\text{E}5\pm 2.\text{E}5$ &\ldots&\ldots& 4 \\
8.3 &\ldots&\ldots&\ldots& $*\,5.\text{E}5\pm 1.\text{E}5$ &\ldots&\ldots& 4 \\
9 &\ldots&\ldots&\ldots&\ldots& $4.\text{E}5\pm 6.\text{E}4$ & $4.2 \text{E}5\pm 6.\text{E}4$ & 5  \\
10 & $2.9\text{E}5\pm 1.\text{E}4$ &\ldots& $1.3 \text{E}5\pm 1.\text{E}4$ &\ldots&\ldots&\ldots& 6 \\
10.02 &\ldots&\ldots&\ldots& $*\,6.\text{E}5\pm 1.\text{E}5$ &\ldots& $7. \text{E}5\pm 3.\text{E}5$ & 7 \\
13 &\ldots&\ldots&\ldots& $*\,1.25\text{E}5\pm 3.\text{E}4$ &\ldots&\ldots& 4 \\
13.1 & $2.5\text{E}5\pm 5.\text{E}3$ &\ldots&\ldots&\ldots&\ldots&\ldots& 8 \\
13.1 &\ldots&\ldots& $8.\text{E}4\pm 2.\text{E}4$ &\ldots&\ldots&\ldots& 9 \\
13.5 &\ldots&\ldots&\ldots&\ldots& $1.17\text{E}5\pm 1.\text{E}4$ &\ldots& 10 \\
14.1 &\ldots&\ldots&\ldots& $*\,1.1\text{E}5\pm 8.\text{E}3$ &\ldots& $1.25 \text{E}5\pm 9.\text{E}3$ & 11     \\
15.6 &\ldots&\ldots&\ldots&\ldots& $9.9\text{E}4\pm 1.\text{E}4$ & $9.8 \text{E}4\pm 1.\text{E}4$ & 5  \\
16.5 &\ldots&\ldots&\ldots& $*\,1.\text{E}5\pm 2.\text{E}4$ &\ldots&\ldots& 4 \\
17.5 &\ldots&\ldots&\ldots&\ldots& $6.6\text{E}4\pm 6.\text{E}3$ &\ldots& 10 \\
18.3 &\ldots&\ldots& $3.75\text{E}4\pm 1.\text{E}4$ &\ldots&\ldots&\ldots& 12       \\
18.3 &\ldots&\ldots&\ldots& $*\,6.\text{E}4\pm 8.\text{E}3$ &\ldots& $7.5 \text{E}4\pm 8.\text{E}3$ & 12 \\
20 &\ldots&\ldots&\ldots&\ldots&\ldots& $5.53\text{E}4\pm 4.\text{E}3$ & 11     \\
22 & $4.45\text{E}4\pm 4.\text{E}3$ & $1.4 \text{E}5\pm 1.\text{E}4$ &\ldots&\ldots&\ldots&\ldots& 14 \\
23 &\ldots&\ldots&\ldots&\ldots& $3.1\text{E}4\pm 5.\text{E}3$ & $3. \text{E}4\pm 5.\text{E}3$ & 5  \\
26 &\ldots&\ldots&\ldots&\ldots& $1.8\text{E}4\pm 1.\text{E}3$ &\ldots& 15 \\
30 &\ldots&\ldots&\ldots& $*\,1.48\text{E}4\pm 1.\text{E}3$ &\ldots& $1.53 \text{E}4\pm 9.\text{E}2$ & 16 \\
30 &\ldots&\ldots&\ldots&\ldots&\ldots& $1.78\text{E}4\pm 1.\text{E}3$ & 11       \\
34.5 & $2.29\text{E}4\pm 1.\text{E}3$ &\ldots& $7.9 \text{E}3\pm 7.\text{E}2$ &\ldots&\ldots&\ldots& 17 \\
38 & $1.3\text{E}4\pm 1.\text{E}3$ & $*\,1.25 \text{E}4\pm 1.\text{E}3$ & $5. \text{E}3\pm 1.\text{E}3$ &\ldots&\ldots&\ldots& 18  \\
38 &\ldots&\ldots&\ldots&\ldots&\ldots&\ldots& 19 \\
38 & $1.32\text{E}4\pm 4.\text{E}2$ & $3.45 \text{E}4\pm 5.\text{E}2$ &\ldots&\ldots&\ldots&\ldots& 20  \\
38 &\ldots&\ldots&\ldots&\ldots& $9.\text{E}3\pm 1.\text{E}3$ &\ldots& 15 \\
45 & $9.14\text{E}3\pm 9.\text{E}2$ & $3.09 \text{E}4\pm 3.\text{E}3$ & $3.31 \text{E}3\pm 3.\text{E}2$ & $3.64 \text{E}3\pm 4.\text{E}2$ & $5.16 \text{E}3\pm 5.\text{E}2$ & $5.58 \text{E}3\pm 6.\text{E}2$ &  48 \\
48.5 &\ldots&\ldots&\ldots&\ldots&\ldots& $5.05\text{E}3\pm 4.\text{E}2$ & 11  \\
55 &\ldots& $1.5\text{E}4\pm 1.\text{E}3$ &\ldots& $3. \text{E}3\pm 1.\text{E}3$ &\ldots&\ldots& 21 \\
81 & $1.8\text{E}3\pm 1.\text{E}2$ &\ldots& $5. \text{E}2\pm 1.\text{E}2$ &\ldots& $1.4 \text{E}3\pm 5.\text{E}1$ &\ldots& 22 \\
81.5 &\ldots&\ldots&\ldots&\ldots& 1120. &\ldots& 15 \\
85 &\ldots& $6.\text{E}3\pm 5.\text{E}2$ &\ldots&\ldots&\ldots&\ldots& 23  \\
85 &\ldots&\ldots&\ldots&\ldots&\ldots& $1.2\text{E}3\pm 8.\text{E}1$ & 11       \\
85 &\ldots&\ldots&\ldots& $9.5\text{E}2\pm 5.\text{E}1$ &\ldots& $1.35 \text{E}3\pm 5.\text{E}1$ & 24        \\
85.7 &\ldots&\ldots&\ldots&\ldots&\ldots&\ldots& 25 \\
100 & $1.13\text{E}3\pm 1.\text{E}2$ &\ldots& $5.25 \text{E}2\pm 3.\text{E}1$ & $4.4 \text{E}2\pm 1.\text{E}2$ & $7. \text{E}2\pm 5.\text{E}1$ & $6.5 \text{E}2\pm 5.\text{E}1$ & 26 \\
100 & $1.\text{E}3\pm 3.\text{E}2$ &\ldots&\ldots&\ldots&\ldots&\ldots& 27 \\
102.5 &\ldots& $5.5\text{E}3\pm 5.\text{E}2$ &\ldots&\ldots&\ldots&\ldots& 28 \\
150 &\ldots& $2.25\text{E}3\pm 3.\text{E}2$ &\ldots&\ldots&\ldots&\ldots& 29 \\
150 &\ldots& $1.8\text{E}3\pm 2\text{E}2$ &\ldots&\ldots&\ldots&\ldots& 23  \\
153 &\ldots& $1.8\text{E}3\pm 2.\text{E}2$ &\ldots& $1.8 \text{E}2\pm 2.\text{E}1$ &\ldots& $2.8 \text{E}2\pm 2.\text{E}1$ & 30 \\
178 &\ldots&\ldots&\ldots&\ldots& $1.7\text{E}2\pm 1.\text{E}1$ &\ldots& 15 \\
178 & $3.3\text{E}2\pm 1.\text{E}1$ & $1.7 \text{E}3\pm 5.\text{E}1$ & $7. \text{E}1\pm 1.\text{E}1$ &\ldots& $1.4 \text{E}2\pm 1.\text{E}1$ &\ldots& 31     \\
200 &\ldots&\ldots&\ldots& $1.04\text{E}2\pm 8.$ &\ldots&\ldots& 32 \\
200 & $2.4\text{E}2\pm 1.\text{E}1$ &\ldots& $1.35 \text{E}2\pm 10.$ &\ldots&\ldots&\ldots& 33 \\
200 & $1.42\text{E}2\pm 4.\text{E}1$ &\ldots&\ldots&\ldots&\ldots&\ldots& 27  \\
250 & $1.4\text{E}2\pm 2.\text{E}1$ &\ldots&\ldots&\ldots& $1.04 \text{E}2\pm 6.$ & $1.07\text{E}2\pm 6.$ & 34 \\
320 & $6.\text{E}1\pm 5.$ &\ldots&\ldots&\ldots&\ldots&\ldots& 35   \\
400 & $4.32\text{E}1\pm 2.$ & $2.24\text{E}2\pm 3.\text{E}1$ &\ldots&\ldots& $2.1 \text{E}1\pm 1.$ &\ldots& 36 \\
404 & $3.6\text{E}1\pm 1.$ &\ldots& $1.65\text{E}1\pm 5.\text{E}{\text{-1}}$ &\ldots& $2.\text{E}1\pm 1.$ &\ldots& 37 \\
408 & $4.3\text{E}1\pm 1.$ &\ldots&\ldots&\ldots&\ldots&\ldots& 38 \\
408 &\ldots& $1.08\text{E}2\pm 2.\text{E}1$ &\ldots&\ldots&\ldots&\ldots& 39 \\
408 &\ldots&\ldots&\ldots&\ldots&\ldots& $1.95\text{E}1\pm 2.$ & 40 \\
600 &\ldots& $8.\text{E}1\pm 5.$ &\ldots&\ldots&\ldots&\ldots& 41 \\
610 &\ldots&\ldots& $6.6\pm 1.\text{E}{\text{-1}}$ &\ldots&\ldots&\ldots& 42 \\
610 &\ldots& $9.\text{E}1\pm 1.\text{E}1$ &\ldots&\ldots&\ldots&\ldots& 43  \\
820 & $8.5\pm 5.\text{E}{\text{-1}}$ & $4.2\text{E}1\pm 6.$ & $5.\pm 5.\text{E}{\text{-1}}$ &\ldots&\ldots&\ldots& 44 \\
1420 & $3.95\pm 5.\text{E}{\text{-2}}$ &\ldots&\ldots&\ldots& $3.4\pm 1.\text{E}{\text{-1}}$ &\ldots& 45 \\
1420 &\ldots& $1.05\text{E}1\pm 5.\text{E}{\text{-1}}$ & $3.3\pm 1.\text{E}{\text{-1}}$ &\ldots&\ldots&\ldots& 46\\
1420 &\ldots&\ldots&\ldots& $3.5\pm 5.\text{E}{\text{-2}}$ &\ldots& $3.6\pm 5.\text{E}{\text{-2}}$ & 47
\end{tabular}
\tablefoot{Columns  names correspond to the positions described in Table \ref{tabla45408raw}. }
\tablebib{(1)~\citet{Ellis1987}; (2) \citet{Hartz1964} ; (3) \citet{Ellis1966};
(4) \citet{Ellis1982}; (5) \citet{Cane1979}; (6) \citet{Caswell1976};
(7) \citet{HamiltonHaynes1968}; (8) \citet{Andrew1966}; (9) \citet{Andrews1969};
(10) \citet{Bridle1967};(11) \citet{Yates1965} (12); \citet{Shain1951}; 
(13) \citet{Shain1954}; (14) \citet{Roger1999}; (15) \citet{Purton1966};
(16) \citet{Mathewson1965}; (17) \citet{Dwarakanath1990}; (18) \citet{Blythe1957};
(19) \citet{Hornby1966}; (20) \citet{Milogradov1973}; (21) \citet{Rohan1970};
(22) \citet{Baldwin1955}; (23) \citet{Wielebinski1968}; (24) \citet{Yates1967};
(25) \citet{Hill1958}; (26) \citet{Bolton1950}; (27) \citet{Piddington1951};
(28) \citet{Ariskin1981}; (29) \citet{Landecker1970}; (30) \citet{Hamilton1969};
(31) \citet{Turtle1962MNRAS}; (32) \citet{Allen1950}; (33) \citet{Droge1956};
(34) \citet{Kraus1957}; (35) \citet{Wall1970}; (36) \citet{Seeger1965};
(37) \citet{Paulini1962}; (38) \citet{Haslam1970}; (39) \citet{Large1961};
(40) \citet{Price1972}; (41) \citet{Piddington1956}; (42) \citet{Howell1970};
(43) \citet{Moran1965}; (44) \citet{Berkhuijsen1972}; (45) \citet{Reich1982};
(46) \citet{Reich1986}; (47) \citet{Reich2001}; (48) This work.}
\end{table*}
 \normalsize

\section{Spectral index map}

From observations at two frequencies $\nu_1$ and $\nu_2$, we can derive
the ``tem\-pe\-ra\-ture spectral index'' $\beta_{\nu_1-\nu_2}$ between
these frequencies, defined as
\begin{equation} \beta_{\nu_1-\nu_2}=-\log\left( \frac{T_{\nu_1}}{T_{\nu_2}} \right)/\log \frac{\nu_1}{\nu_2},
\label{defie}\end{equation}
where $T_{\nu_1}$ and $T_{\nu_2}$ are the respective brightness
temperatures.

At 45 and 408 MHz, the relation between the brightness temperature and
the intensity of the radiation approximately follows  the
Rayleigh-Jeans law,  which implies that it is a simple relation between the temperature spectral
 index $\beta$ and the intensity spectral index $\alpha$ given by 
 $\beta=\alpha + 2 $.

To calculate these  spectral indices appropriately it is necessary to use
brightness temperatures measured at the same resolution. Because of
the difference in beam-size (and shape) between the surveys at 45 and
408 MHz, we  convolved each map to $5^\circ$ FWHM.

 The measured temperature $T_\nu$ at frequency $\nu$ is modeled as
\begin{equation} 
\label{ecuaciongeneralTT}T_{\nu}=T_{\nu,G}+T_{\nu,0}=T_{\nu,G}+T_{CMB}+T_{\nu,Ex}+T_{\nu,ZLC}  
\end{equation}
with a Galactic component $T_{\nu,G}$ and an isotropic one $T_{\nu,0}$
(the ``offset''), which includes the non-thermal extragalactic {background}
($T_{Ex}$), the CMB ($T_{CMB}$), and a zero-level correction
($T_{ZLC}$). To obtain a spectral index associated with
Galactic emission, we need to correct the observed temperature by
subtracting the offset temperature. 
If the spectral index
$\beta$ due to Galactic emission is constant in some area, then the
following relations hold:
\begin{equation}\label{ecuaciongeneral2TT}  
\begin{split}
T_{\nu,G} &=(T_\nu-T_{\nu,0})\propto\nu^{-\beta}, \\
\frac{T_{\nu_1}-T_{\nu_1,0}}{T_{\nu_2}-T_{\nu_2,0}} &=\left(\frac{\nu_1}{\nu_2}\right)^{-\beta}. 
\end{split}
\end{equation}

\subsection{Zero-level correction of the 45 and 408 MHz surveys}
On the basis of published data at different frequencies, we constructed a
multi-frequency spectrum at selected and well-observed positions of
the sky. We use these spectra to determine  the zero-level correction  temperature  for the  45 and 408 MHz surveys. { This zero level is a second order correction to the already calibrated
data, and the method presented here is  an  alternative to the more common  approach
of using TT-plots \citep[see e.g.][]{Reichmap1988}.}
\\ The six selected positions { are}:
\begin{itemize}
\item{The Galactic poles and the anticenter: these regions were
chosen because we expect a minimal amount of  Galactic thermal absorption in these
directions.}
\item{ The directions of minimum temperature at 45 MHz in the northern
($l=192^\circ\, ,\, b=48^\circ $) and southern ($ l=235^\circ\, ,\,
b=-45^\circ $) Galactic hemispheres:  these directions were selected
because the spectral index calculated in these zones is very sensitive
to any correction applied to the data.}
\item{The ``calibration point'' ($l=38^\circ\, ,\, b=-1^\circ $):
so-called because it was used to calibrate the \mbox{45 MHz} southern
survey \citep{45south}.  }
\end{itemize}

Multi-frequency spectral data corresponding to these six positions are
given in Table \ref{tablaseiszonas}.  
Whenever these values were not 
in digital format in the literature,
they were read directly
from the published contour maps.  If the publication 
 does not provide an estimate of the errors, we use 
plus-minus half of the difference between 
 the closest contours.   

We  applied two extragalactic corrections $T_{CMB}$ and $T_{\nu,Ex}$ (see Eq. \ref{ecuaciongeneralTT}):
\begin{itemize}
\item{The 2.7 K of the CMB represents different percentages of the
sky brightness temperature depending on the position, but especially on
the frequency $\nu$ of the radiation. While at frequencies such as
\mbox{45 MHz}, where the minimum temperature is close to 3500 K, this
correction is negligible, it is not at \mbox{408 MHz}, where the
minimum gets close to 13 K.}

\item{The extragalactic non-thermal background. This component is
believed to originate from the integrated emission of non-resolved
extragalactic radio sources. Appendix \ref{extra} gives a { robust}
estimate of this
extragalactic non-thermal spectrum (ENTS, Fig. \ref{ENTS}).
 We subtract this
spectrum from every multifrequency plot , in addition to the 45 and 408 MHz
data. This extragalactic emission corresponds to  $\sim1100$ K at
45 MHz and $\sim2.4$ K at 408 MHz.  }
\end{itemize}

\begin{table}
\centering
\caption{Uncorrected temperatures and spectral index.} 
\label{tabla45408raw}
\begin{tabular}{lcccc} \hline\hline
Zone & Position & T$_{45}$ &T$_{408}$ & $\beta_{45-408}$ \\
&&~K&K& \\\hline
Min. South&l=235$^\circ$, b=-45$^\circ$&3.64E3&1.41E1&2.52 \\
Min. North&l=192$^\circ$,b=48$^\circ$&3.31E3&1.33E1&2.50\\
Cal. Point&l=38$^\circ$,b=-1$^\circ$&3.09E4&1.96E2&2.30\\
SGP&b=-90$^\circ$&5.58E3&2.04E1&2.55\\
NGP&b=90$^\circ$&5.16E3&1.92E1&2.54\\
Anticenter&l=180$^\circ$,b=0$^\circ$&9.14E3&3.85E1&2.48\\\hline
\end{tabular}
\end{table}

\begin{figure}
   \centering
   \includegraphics[width= 0.55 \textwidth, angle=-90]{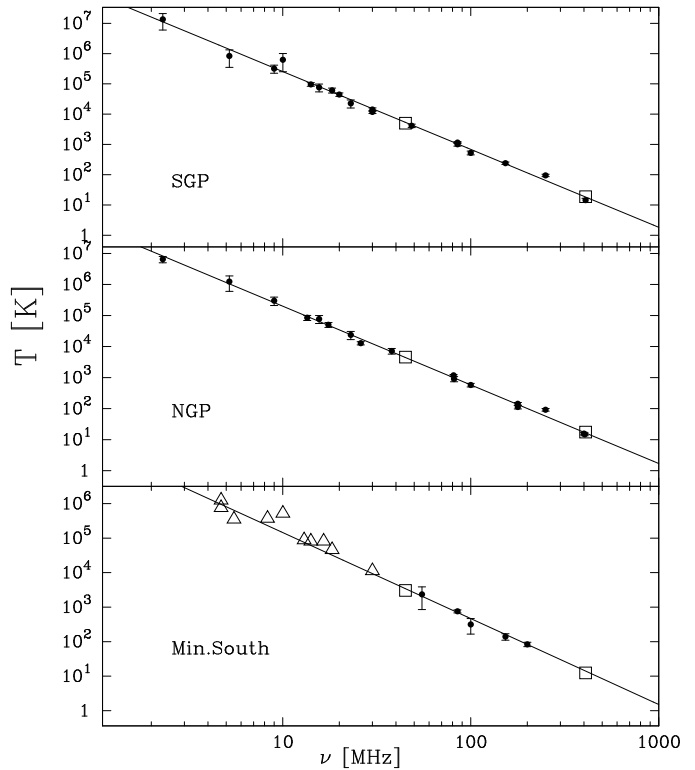}
   \caption{Multi-frequency spectra of the  Galactic component for (from top to bottom): South Galactic Pole, North Galactic Pole, and minimum-south zones. The triangles are not considered in the least squares fitting of the line. The squares represent the 45 and 408 MHz data.}\label{multifrecuencias-surpolos}
 \end{figure}

\begin{figure}
   \centering
   \includegraphics[width= 0.55 \textwidth, angle=-90]{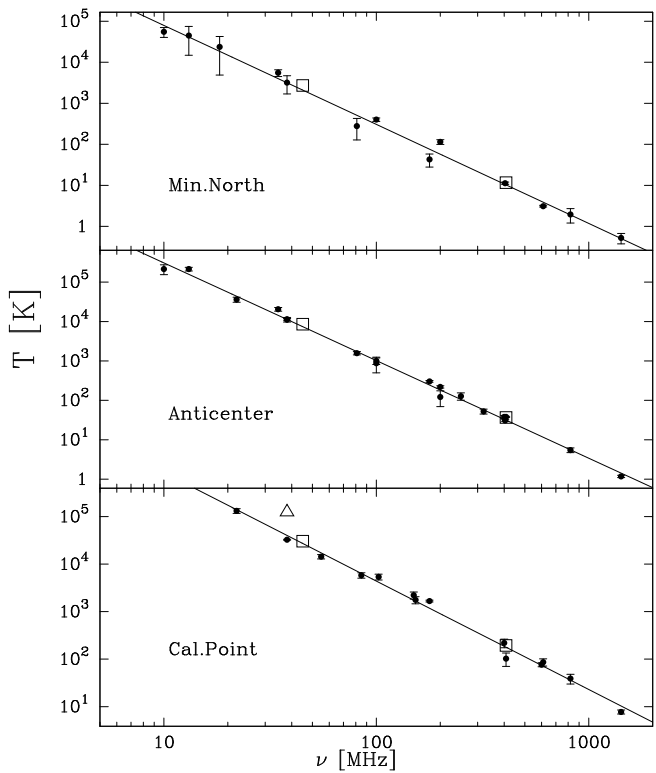}
   \caption{Multi-frequency spectra of the Galactic component for the (from top to bottom)  minimum-north, calibration point and anticenter zones. The triangles are not considered on the least squares fitting of the line. The squares represent the 45 and 408 MHz data.}\label{multifrecuencias-norcaliban}
 \end{figure}

\begin{figure}
   \centering
   \includegraphics[width= 0.5 \textwidth]{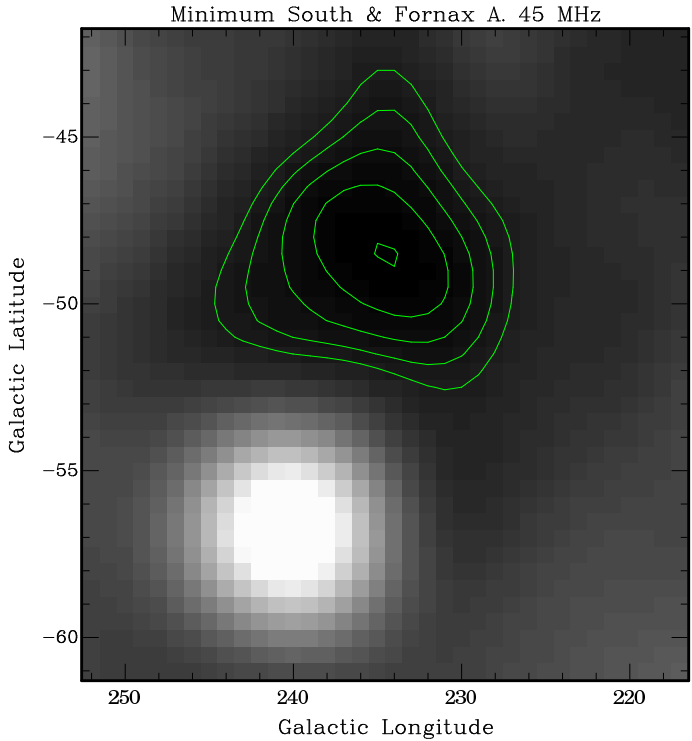}
   \caption{Minimum-South zone, Plate-Care\'e projection and five equi-spaced contours between 3615 and 3870 K. No extragalactic 
     correction applied. The source seen next to the contours is Fornax A.}\label{minsouthzone}
 \end{figure}

\begin{table*}
\centering
\caption{ Extragalactic corrected temperatures and spectral indices.}
\label{tablaextracorr}
\begin{tabular}{lccccc} \hline\hline
Zone &T$_{45}$ &T$_{408}$ & $\beta_{45-408}^{extra\_corr}$&$\beta^{extra\_corr}_\text{multifreq}$&$\Delta\beta$\\
 &K&K& & & \\\hline
Min. South & 2.51E3 &   8.98 & 2.55 & 2.49 & -6.3E-2\\
Min. North & 2.18E3 &   8.18 & 2.53 & 2.41 & -12 E-2\\
Cal. Point   & 2.98E4 & 1.91E2 & 2.29 & 2.28 & -0.9E-2\\
SGP          & 4.45E3 & 1.53E1 & 2.57 & 2.57 & -0.0E-2\\
NGP          & 4.03E3 & 1.41E1 & 2.57 & 2.53 & -3.1E-2\\
Anticenter   & 8.01E3 & 3.34E1 & 2.49 & 2.48 & -0.7E-2\\\hline
\end{tabular}
\end{table*}
Figures \ref{multifrecuencias-surpolos} and
\ref{multifrecuencias-norcaliban} show the spectra of the six chosen zones
together with least squares (in the log-log scale) fitted lines. Both
extragalactic components have been subtracted from the data, and the
defined multi-frequency spectral index is the slope of this line.  In Table
\ref{tablaseiszonas}, the * symbol indicates that the point is not included
in the least squares fitting. We neglected as an outlier one data value
from the calibration point multifrequency fit
(Fig. \ref{multifrecuencias-norcaliban}, bottom panel).  This value,
corresponding to 38 MHz \citep{Blythe1957}, lies above the fit by a factor
of $\sim3$. We also neglected the data from the minimum south point below
45 MHz, for reasons we detailed below. Neglected data are marked by
triangles in Figs. \ref{multifrecuencias-surpolos} and
\ref{multifrecuencias-norcaliban}, where the 45 and 408 MHz data are
identified by squares.  Large discrepancies could be due to real physical
features such as thermal absorption at very low frequencies, the steepening
of the spectral index towards higher frequencies, instrumental calibration
errors, etc \ldots The points in the minimum-south plot
(Fig. \ref{multifrecuencias-surpolos}, bottom panel) below 45 MHz were also
ignored because we believe they do not represent the real emission of this
zone as the large beams associated with these surveys (from 7\degs to
15\degs~ FWHM) wash-out the relatively small area of the
minimum-south. Figure \ref{minsouthzone} illustrates this effect: the
contours delineate the minimum south, and the unresolved source just at the
side seen in grey-scale is Fornax A, which also represents the 45 MHz
antenna beam size.  A larger beam would confuse the flux that surrounds the
minimum zone (Fornax A flux inclusive) and would infer a higher temperature
in this direction. This interpretation is confirmed in Fig.
\ref{multifrecuencias-surpolos} where we can see that the fitted line to the
data at frequencies {\it above} 45 MHz lies below of most of the data points
measured at lower frequencies.  We tested the spectra of the other five
selected areas, and found that removing the low-frequency data does not
affect the quality of the fittings significantly. In particular, the
minimum-north zone is far more extended than the minimum-south zone, so the
antenna beam size does not affect the {measured temperature}.

 In Figs. \ref{multifrecuencias-surpolos} 
and \ref{multifrecuencias-norcaliban}, we did not include the
points at 45 and \mbox{408 MHz} in the fit  because we are trying to find a
correction applicable to these surveys based on independent data.

The coordinates of the six zones, the uncorrected temperatures at 45 and
408 MHz, and the spectral index derived from these two values are given in
Table  \ref{tabla45408raw}. In contrast, Table  \ref{tablaextracorr}
shows temperature and spectral index values corrected by the extragalactic
extended components, that is, the ENTS and the CMB. The columns of
Table \ref{tablaextracorr}\ are: ``extragalactic-corrected''
tem\-pe\-ra\-tu\-res, the ``extragalactic-corrected'' spectral index
derived between 45 and 408 MHz, the ``extragalactic-corrected''
multifrequency spectral index, and the difference between these last two
columns
($\Delta\beta=\beta_\text{multifreq}^{extra\_corr}-\beta_{45-408}^{extra\_corr}$).

 We now explain  the quantitative criterion used to evaluate the
 quality of the fits: having seen that extragalactic-corrected
 multi-frequency spectra are well fitted by single power laws, as
 expected if we observe non-thermal synchrotron radiation, we
 determine  zero-level corrections (ZLC) to the 45 and 408 MHz maps
 such that, when applied uniformly to the six zones, they maximize the
 sum of the squares of the linear correlation coefficients of the six
 linear fits.
 \begin{table}
\centering 
\caption{Temperature offset corrections in K. Positive values should
  be subtracted, while negative should be added according to equation
  \eqref{ecuaciongeneralTT}.}\label{corrections}
\begin{tabular}{rcc}\hline\hline
Component & 45 MHz & 408 MHz \\\hline
CMB & 2.7 & 2.7 \\
ENTS\tablefootmark{a} & 1100 & 2.4\\
ZLC\tablefootmark{b} &  -544 &  -3.46 \\\hline
\end{tabular}
\tablefoot{\tablefoottext{a}{Extragalactic non-thermal spectrum.}
\tablefoottext{b}{Zero-level correction (see Eq. \ref{ecuaciongeneralTT}).}}
\end{table}
 
To summarize, reliable Galactic spectral indices
 between the 45 and \mbox{408 MHz} can be derived  by applying 
the ENTS, CMB and
 the ZLC corrections to both surveys.  Table
 \ref{corrections} displays all of these corrections. 
Table  \ref{tablacorrected} shows the 45-408 MHz spectral index  corrected values (corrected for ENTS, CMB, and ZLC)  and the differences between these corrected temperatures  and the multifrequency fit ($T_{\text{fit}}$).
   These spectral
 indices and the map derived represent  our best estimates of the
 \emph{Galactic} temperature spectral index between these two
 frequencies.

\begin{figure*}
\centering
\includegraphics[angle=90, width= .7 \textwidth]{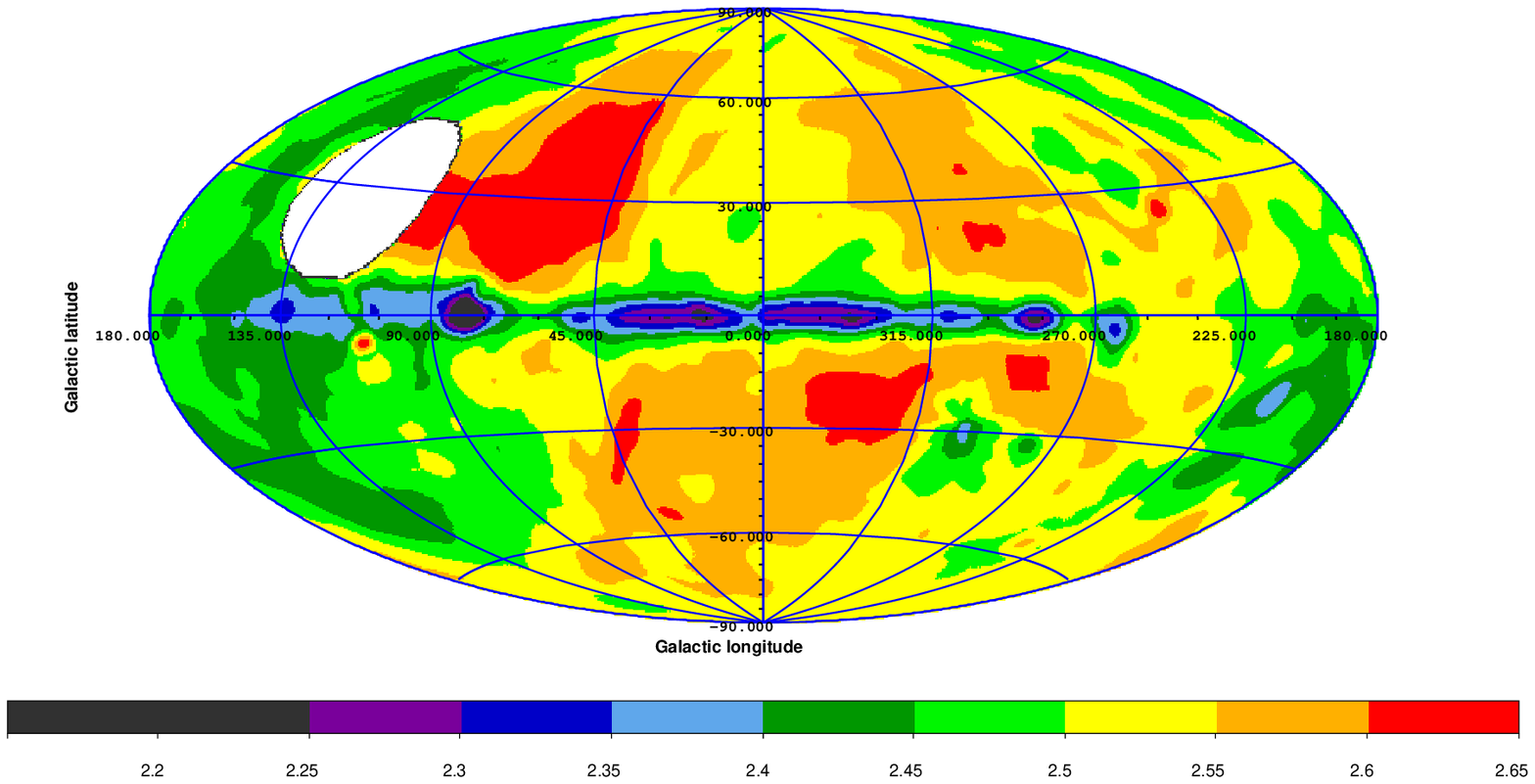}%
\caption{Galactic temperature  spectral index between 45 and \mbox{408 MHz} with the corrections of Table \ref{tablacorrected} applied. The map has $5^\circ\times5^\circ$ resolution.}\label{iecorrmult}
\end{figure*}

The isotropic corrections
 ($T_{\nu,0}=T_{CMB}+T_{\nu,Ex}+T_{\nu,ZLC}$ in
 Eq. \ref{ecuaciongeneralTT}) are
\begin{equation}T_{\text{45},0}=550\text{  K \,\,\,\,\,and  }\quad T_{\text{408},0}=1.6\text{  K}. \end{equation}
Figure \ref{iecorrmult} shows the final spectral index map.

\begin{table}
\caption{Corrected 45-408 MHz spectral indices and differences between the corrected  temperatures  and the multifrequency fit. 
}\label{tablacorrected}
\centering
\begin{tabular}{lcccc}\hline\hline
Zone    & $\beta_{45-408}^{uncorr.}$ & $\beta_{45-408}^{corr.}$  &   $\frac{T_{45}-T_\text{fit}}{T_{45}}$\%%
  &  $ \frac{T_{408}-T_\text{fit}}{T_{408}}$\% \\\hline
Min. South & 2.52 & 2.50 &-10  & -12 \\
Min. North & 2.50 & 2.47 & 18  & 10  \\
Cal. Point   & 2.30 & 2.29 & 10  & 8.5 \\
SGP          & 2.55 & 2.53 &-6.1 & 2.1\\
NGP          & 2.54 & 2.52 & 3.3 & 5.5\\
Anticenter   & 2.48 & 2.47 & 12  & 14  \\\hline
\end{tabular}

\end{table}

 There are several considerations to take into account concerning the
multifrequency spectrum procedure:
 \begin{itemize}
 \item{ The different  resolutions  of the data. When comparing 
different surveys, we really should adopt a common
   resolution,  which we do not here. However, apart from
   the systematic errors in the measured temperatures at low frequencies in the
   minimum-south zone, as already explained, the multifrequency spectra show remarkable
   agreement between independent measures and no significant departure
   from a single power-law fit.  Excluding
     the minimum-south zone, the zones selected correspond to
     relatively extended and uniform areas at the resolution of
     the studies we used. Therefore, differences in beam sizes should
     not affect significantly  the temperature measured in these directions. }

 \item{Free-free absorption should produce departures from a
   single power law, specially at low frequencies that becomes significant at the ``calibration point'', which is close to the
   Galactic plane towards the inner Galaxy. However, the spectrum shown in Fig. \ref{multifrecuencias-norcaliban} (bottom panel)
   does not show any systematic departure from the straight line fit,
   at least one that is noticeable beyond  the scatter of the points. 
Therefore, a single
   power-law seems to be a reasonable fit in each zone.}
 \item{Apart from the intrinsic shortcomings of the multifrequency data
caused by  their low resolution, we believe that these data are very
consistent and, in general, we did not find calibration or
instrumental errors that  question their quality.   The determination
of spectral indices by means of  multifrequency power-law fits should
be statistically stronger.}
 \end{itemize}




\section{Discussion of the maps}

\subsection{Scanning and matching effects}

 The data from low-frequency Galactic surveys are  usually obtained
by scanning the sky with transit instruments.  By ``scanning
effects'', we refer to the stripes produced in the temperature maps
(and particularly noticeable in the spectral index map) because of  the
different observing conditions, e.g. differences in gain or
ionospheric absorption, between adjacent scans made by the instrument.
Another spurious effect is produced when surveys made at
different epochs or zones or with different instruments are combined in a
process described, for example, in \citet{Haslam1981} where  the 408
MHz all-sky map was obtained. This effect is caused by the matching of
 partial maps that together form the final map, and we refer to this as the
 ``matching effect''.  All these spurious features and residues are
usually produced along constant declination and/or right ascension.\\
The 45 MHz map has a matching effect produced by combining the
northern and southern surveys at $\delta\approx10^\circ$.  {
\citet{45north} calibrated the northern 45 MHz survey using data of
the southern survey in a strip of the sky observed with both
instruments.  The same procedure was used to create the
all-sky 45 MHz map presented in this paper}. Scanning effects in this
survey are present especially at the positions  RA: $15h\lesssim RA \lesssim 24h$
and $-30^\circ\lesssim DEC\lesssim 30^\circ$.  They appear at constant
declination because both radio telescopes used in the 45 MHz survey
were transit instruments. Stripes in the spectral index map at
constant RA are due to scanning and matching effects associated with the
408 MHz map, which is formed by combining six  different surveys
as described in \citet[Fig 1]{Haslam1981}. Both effects are more
important near the zones where the temperature reaches a minimum and
are far more  noticeable in the spectral index map rather than in the
temperature map.  The effects described produce some systematic errors
in our spectral index at a level lower than 0.05.

\subsection{ Principal physical processes and characteristics}
 The principal emission mechanism at 45 MHz is synchrotron radiation from
 relativistic electrons that permeates the interstellar medium (ISM). The
 emissivity at a frequency $\nu$ is proportional to
 $H^\frac{\epsilon+1}{2}\nu^\frac{\epsilon-1}{2}$, where $(-\epsilon)$ is
 the exponent of an assumed power-law distribution of the relativistic
 electrons energy and H is the magnitude of the magnetic field.  A study of
 synchrotron emission alone cannot differentiate between the effect of the
 magnetic field and the characteristics of the fast electron
 population. Using pulsar rotation measures, \citet{Han2006} placed
 constraints on the global magnetic field ($2\sim3\,\mu$G on average in the
 ISM).  \citet{Testori2008} also note from polarization maps of the 21cm
 continuum that the polarized radio emission is only a few percent (5\%) of
 the total radiation.  This may imply that on Galactic scales the
 ``random'' component of the magnetic field is far more important than
 coherent field zones.  If this were the case, synchrotron emission would
 give a direct way of obtaining the relativistic electron column
 density. If, however, as the three-dimensional Galactic models of
 \citet{Sun2008} suggest, the regular field is comparable and even more
 intense than the random field, then the relativistic electron column
 density can only be obtained from the synchrotron emission with the
 knowledge of the direction of the regular magnetic field.

\begin{figure}
\centering
\includegraphics[width= 0.46\textwidth]{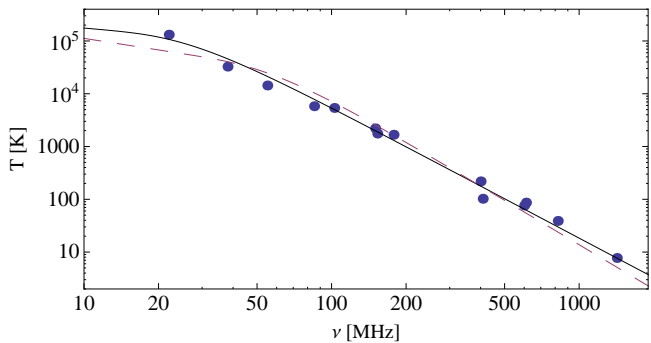}
\caption{Fittings to the calibration-point data using a thermally absorbed
power-law spectrum with spectral index equal to that at the anticenter (2.48, continuous line) and  the value used by \citet{Jones1974} (2.80, dashed). Both fittings
are reasonable considering the dispersion in the data. }\label{absorbedcp}
\end{figure}

The principal absorption mechanism at 45 MHz is thermal absorption from
warm ($\sim 8000$ K) ionized hydrogen.  This absorption diminishes the
spectral index in the Galactic plane with respect to that at higher
latitude. In the multifrequency data, we have seen no obvious absorption
feature or departures from a straight-line fit not even in the
calibration-point zone (spectral index 2.28), where it would be expected
because it is closer to the Galactic plane and in the direction of the
inner Galaxy.  An absorbed spectra may not differ very much from a
non-absorbed one across this range of frequencies. Figure \ref{absorbedcp}
presents two fits to the calibration-point data assuming that the
absorption occurs homogeneously along the line of sight. The one
  displayed with a continuous line assumes that the non-absorbed, purely
emission spectrum has an spectral index of 2.48 (the anticenter value),
while the other (dashed line), uses a steeper spectrum with spectral index
of 2.8.  The values derived for
$\text{EM}/(\text{cm}^{-6}\text{pc})\times(T_e/1000\text{K})^{-1.35}$,
where EM is the emission measure and $T_e$ is the electron temperature, are
482 and 455, very close to the 470 obtained by \citet{Jones1974} towards
the $l=-35^\circ, b=0^\circ$ direction, which is located in an
approximately symmetrical longitude about the Galactic center with respect
to the calibration-point ($l=38^\circ, b=1^\circ$).  Both fittings are
fairly reasonable, and the absorption column can take account of the
flattening of the spectrum.  The emission measures derived assuming
$5000\,{\rm K}<T_e<8000\,{\rm K}$ for the warm ionized gas to produce the absorption seen in
the spectral index map are between 1600 and 4000 pc cm$^{-6}$, well above
the emission measures from the warm ionized medium \citep{Hill2008}. We
conclude that this absorption is produced mainly in HII-regions, unresolved
by the present survey, implying that the emission measures derived should
be regarded as beam-averaged values.

An important characteristic of the map is the presence of two minimum
temperature zones, whose multifrequency spectra are shown in
Figs. \ref{multifrecuencias-surpolos} and
\ref{multifrecuencias-norcaliban}, in the bottom and top panels,
respectively.  They do not exhibit absorption in the spectral index map
discernible by means of a flatter spectrum or a turn-over towards lower
frequencies in the multifrequency data.  The two zones could be a single
volume with either a low density of relativistic electrons or weak magnetic
fields. This volume should be nearby judging from its large extension in
the sky. There appears to be little correlation with the local bubble
described in \citet{Lallement2003}, but the positions of the minima are
consistent with the super bubble found by \citet{Heiles1998}.  Interpreting
the minimum zones as being produced by a bubble is consistent with both the
little absorption detected in their directions and the interpretation
of \citet{Sun2008} that high-latitude radio emission comes not from an
extended halo-zone but primarily from a local excess of synchrotron
emissivity: a local bubble could then explain the minima in these
directions.  The polarization map of \citet{Testori2008} does not show any
conspicuous enhancement or decrement in  the polarization fraction
toward the southern minimum. We also note that the temperatures measured
towards the minima can be considered as upper limits to the extragalactic
background.

The maximum spectral index ($\sim$2.7) is found in a rather flat
plateau of approximately 15\degs diameter around $l\sim70^\circ$,
$b\sim+25^\circ$ { adjacent to the Northern Spur}. This feature is consistent with the results obtained
by \citet{Webster1974} (at $\delta=40^\circ$) and \citet{Lawson1987}
who found a steepening of the spectrum in this zone, that is, on the
ridge of the Northern Spur rather than on the spur itself. This area
of high spectral index is also noted by \citet{Roger1999} 
between 22 and 408 MHz.\\
The mean values of the spectral index are
consistent with studies made at similar frequencies, for example,
\citet{Purton1966} found an spectral index in the halo-high latitude
zone of 2.57, close to our mean of 2.54 and 
\citet{Sironi1974} found a spectral index of 2.53
in the minimum-north between 81.5 and 408 MHz (our  value is 2.47).   
Agreement is better if the zones compared have a relatively constant spectral index because in this case the difference in beam sizes does not affect the spectral index. In contrast, when looking at the anticenter, \citeauthor{Purton1966} gets
2.55, a higher value than ours, but within our errors.

The all-sky map compiled by
\citet{Reichmap1988} measures  spectral indices that are systematically 
higher by
about  0.2 than our map. \citet{Webster1974}, who
measured the spectral index between 408 and 610 MHz, also obtained values
slightly larger than ours. In general, the spectral index appears to
 increase slightly with frequency in every zone,  probably because of 
aging of the relativistic electron population.
The increase in spectral index with
frequency on Galactic scales is also obtained by \citet{Oliveira2008} through their negative
$\gamma$-parameter at least at $\nu\lesssim1$ GHz, where synchrotron 
dominates over free-free or the CMB.  
  We do not see a flattening of the
spectrum with increasing Galactic latitude at $l\sim180^\circ$ as
described in \citet{Reich1988}.

\section{Conclusions}
We have presented a 45 MHz survey that covers more than $ 95\%$ of the
sky and has the advantage of being assembled by data taken with
  only two radio telescopes of similar characteristics. This map shows the
emission from the relativistic electrons being decelerated by the magnetic
fields in the interstellar medium. There are two zones of minimum
temperature toward $l\sim230^\circ$ that are located more or less
symmetrically with respect to the Galactic plane.  They presumably belong
to a single local decrement in the density of relativistic electrons or of
the magnetic field magnitude.\\ We have used this map together with the 408
MHz all-sky survey from \citet{Haslam1981} to produce an all-sky spectral
index map. We have used a large literature compilation of data to: i)
estimate the extragalactic non-thermal emission contribution and ii)
implement a method for finding zero-level corrections for radio surveys
based on independent data.\\ The spectral index map has allowed us to
derive some conclusions about overall Galactic features: the sky has
spectral indices that range between $2.1$ and $2.7$.  Over most of the sky,
the index is between $2.5$ and $2.6$, which is reduced by thermal
absorption to values between $2.1$ to $2.5$ across the Galactic plane strip
($|b|<10^\circ$). This absorption is probably due to HII regions rather
than a more extended global structure such as the WIM.  There is a large
zone around $l\sim70^\circ$, $b\sim+25^\circ$ adjacent to the Northern
Polar Spur where the average spectral index is close to the maximum of
$2.7$.\\ We have also reviewed (Appendix \ref{extra}) estimates of the
extragalactic non-thermal contribution in the low frequency range.

\begin{appendix} 
\section{The extragalactic background spectrum}\label{extra}
The knowledge of the extragalactic non-thermal spectrum (ENTS),
 besides its importance on its own, permits a more accurate determination 
of the emission of our own Galaxy and  is another ``foreground''
 component for CMB studies. To { separate} the radio emission of our
 own Galaxy from this extragalactic {background} emission, we have searched the
 literature to  find a reliable estimate of the ENTS.  Table \ref{tablextragal}
 displays the works selected for their \emph{original} estimate of
 the extragalactic background component.
We note that although this  is
 considered an important issue in most radio surveys, only a few
 authors have made attempts to estimate it:\\ 
i) The most common approach is to consider 
 the ENTS as the integrated contribution of extragalactic radio
 sources, extrapolating their population and distribution from
 catalogs, and using a cosmological model.
 \citet{Bridle1967,Simon1977,Lawson1987}, and \citet{Gervasi2008} 
 used and described the method in detail. In general, the ENTS derived
 has a spectral index that is about $0.1$-$0.2$ higher than the average emission
 from our own Galaxy or any other ``normal'' galaxy. This is
 attributed probably to the influence of the scarcer but much more
 luminous radio-galaxies { on} the ENTS. \\ ii) Another  less common
 approach is to model the Galactic emission to extract it
 from the measured data. \citet{Yates1968} and \citet{Cane1979} 
 used simple models consisting of axisymmetric cylindrical components
 plus arms, spurs, and Galactic sources.\\ iii) \citet{Shain1959} used
 30-Doradus as a thermal absorption screen  that at low frequencies 
 blocks the ENTS. Comparison of the emission measured along its line
 of sight with a nearby one,  not-blocked by this HII-region, would
 cancel out the Galactic component and leave the extragalactic emission.
\\ iv)
 \citet{Baldwin1967} uses differential spectra  to cancel the
 ENTS and obtain tighter constraints on the Galactic spectrum, which,
 together with an adequate treatment of the instrumental components,
 would permit us to obtain the ENTS.

\begin{figure}
\centering
\includegraphics[width= 0.45 \textwidth]{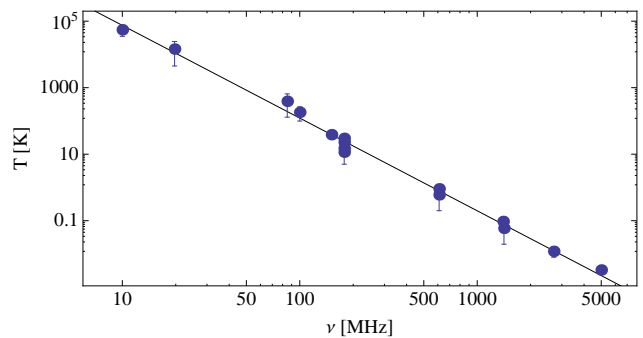}
\caption{Extragalactic component estimates and a single power-law fit.}\label{ENTS}
\end{figure}

 Figure \ref{ENTS} shows a single power-law fit to the
 temperatures shown in Table \ref{tablextragal}.  
The best-fit solution is given by
 \begin{equation}
\log T=7.66-2.79  \log\left(\nu/\text{MHz}\right).
\label{fitENTS}
\end{equation}

Whenever the publication was not explicit about the uncertainty 
in the derived value, we assume it to be twice as large as  the errors
  in the data from where 
the extragalactic component was derived. 
 From the uncertainty in the data points, we
 estimate the error in the derived ENTS spectral index to be $\gtrsim 0.04$.
 \begin{table}
 \centering
\caption{Extragalactic data obtained from the literature. Lower and upper limits to the ENTS.}\label{tablextragal}
\begin{tabular}{rccl}\hline\hline
$\nu$ & Lower & Upper & Ref. \\
MHz&K&K&\\\hline
10 & 4E4 & 7E4 & 1 \\
19.7 & 7E3 & 2.2E4 & 2 \\
85 & 195. & 585. &  3\\
100 & 120 & 244 & 4 \\
151 & 36.3 & 40.9 & 5 \\
178 & 7.5 & 22.5 & 6 \\
178 & 10 & 13 &  7\\
178 & 19.8 & 25.8 & 5 \\
178 & 23 & 37 & 8 \\
610 & 0.3 & 0.9 & 9 \\
610 & 0.84 & 0.98 & 5 \\
1400 & 0.088 & 0.102 & 5 \\
1415 & 2.95E-2 & 8.8E-2 & 9 \\
2700 & 9.1E-3 & 1.5E-2 & 5 \\
5000 & 3.E-03 & 3.6E-3 & 5 \\
8440 & 3.5E-4 & 5.6E-4 & 5 \\\hline
 \end{tabular}
\tablebib{(1)~\citet{Cane1979}; (2) \citet{Shain1959}; (3) \citet{Yates1968};
(4) \citet{Clark1970}; (5) \citet{Gervasi2008}; (6) \citet{Veron1967};
(7) \citet{Simon1977}; (8) \citet{Bridle1967}; (9) \citet{Lawson1987}}
  \end{table}
 
\end{appendix} 

\begin{acknowledgements}
We would like to thank Dr. Patricia Reich for producing the 
 $5^\circ$ degraded resolution maps at  45 and 408 MHz. J.M. acknowledges partial support from Centro de Astrof\'isica FONDAP 15010003 and from Center of Excellence in Astrophysics and Associated Technologies (PFB 06).
\end{acknowledgements}

\bibliographystyle{aa}
\bibliography{bibliografia}

\end{document}